\documentclass{ws-procs975x65}

\newcommand{\bsig}{\hbox{\boldmath $\sigma$}}
\newcommand{\btheta}{\hbox{\boldmath $\theta$}}

\newcommand{\citenumber}[1]{[\citenum{#1}]}

\begin{document}

\markboth{D. Hestenes}{Gauge Gravity and Electroweak Theory}

\wstoc{Gauge Gravity and Electroweak Theory}{D. Hestenes}

\title{GAUGE GRAVITY and ELECTROWEAK THEORY}

\author{DAVID HESTENES}

\address{Physics Department, Arizona State University,\\
Tempe, Arizona 85287--1504, USA\\
\email{hestenes@asu.edu}}

\begin{abstract}
Reformulation of the Dirac equation in terms of the real Spacetime Algebra
 (STA) reveals hidden geometric structure, including a geometric role for
the unit imaginary as generator of rotations in a spacelike plane. The
STA and the real Dirac equation play essential roles
in a new Gauge Theory Gravity (GTG) version of General Relativity (GR).
Besides clarifying the conceptual foundations of GR and facilitating complex
computations, GTG opens up new possibilities for a unified gauge theory of
gravity and quantum mechanics, including spacetime geometry of electroweak
interactions. The Weinberg-Salam model fits perfectly into this geometric
framework, and a promising variant that replaces chiral states with Majorana
states is formulated to incorporate zitterbewegung in electron states.
\end{abstract}

\bodymatter

\section{Introduction}\label{Intro}

This conference rightly honors Marcel Grossman as exemplar of the
profound influence of mathematics in shaping physical theory. He
educated Einstein in the {\it absolute differential calculus} of Ricci
 and Levi-Civita and extended the concept of {\it tensor} to covariant systems.
As we all know, this calculus provided Einstein with new conceptual structures
that were essential for creating his General Theory of Relativity.

The conference parallel session on {\it Geometric Calculus and Gauge
Theory Gravity} presented a more recent approach to gravity theory where
alternative mathematics plays an essential role. Though few physicists
are yet conversant with this approach, it is developing vigorously,
with a substantial body of new mathematical techniques and physical
results. This article is a status report on the field with emphasis
on recent developments and prospects for an integrated gauge theory
of gravity and electroweak interactions. Readers are enjoined to
reconsider many things they already know from the new perspective
of {\it Geometric Calculus!}

\section{Historical roots and recent progress}\label{Back}

Geometric Calculus (GC) is an extension of Geometric Algebra (GA)
to include differential and integral calculus on manifolds. From a
practical point of view, it can be regarded as an extension of tensor
algebra and calculus to integrate spinors and differential forms in a
unified, coordinate-free mathematical system. GA has its roots in the
work of Grassmann and Clifford in the middle of the nineteenth century,
but its development as a unified mathematical language for physics did not
 begin until 1966 \cite{Hest66}. Details of this development are reviewed in
{\citenumber{Hest03a, Hest03b, Hest05}}, and a comprehensive
treatment is given in {\citenumber{Doran03a}}. To date,
there are more than 10 books and 200 articles on GA and its applications
to physics, mathematics, engineering and computer science. Most of the
GA literature can be traced via websites \cite{GC URL}, from which all the
articles cited in this report can be downloaded.

The specialization of GA to Minkowski spacetime is called {\it Spacetime
Algebra} (STA). As explained below, STA clarifies the geometric
significance of the Dirac Algebra and thereby extends its range of
application to the whole of physics. In particular, STA provides the
essential mathematical framework for the new {\it Gauge Theory Gravity} (GTG).

The foundations of GTG are fully expounded in a seminal paper {\cite{Lasenby98}},
so we can concentrate on highlights of its unique features. GTG is a gauge
theory on Minkowski spacetime, but locally it is equivalent to {\it General
Relativity} (GR), so it can be regarded as an alternative formulation of
GR.\cite{Heyl76} However, GTG reformulates (or one might say, replaces) Einstein's
 vague principles of equivalence and general relativity with sharp gauge
principles that have clear physical consequences (Section IV). These
gauge principles are more than mere rephrasing of Einstein's ideas. They
lead to {\it intrinsic} mathematical methods that simplify modeling and
calculation in GR and clarify physical meaning of terms at every stage.
In particular, they provide clean separation between gauge transformations
and coordinate transformations, thus resolving a point of longstanding
confusion in GR. Moreover, GTG simplifies and clarifies the analysis of
singularities, for example, in assignment of time direction to a black
hole horizon. Finally, since tensors and spinors are fully integrated
in STA, GTG unifies classical GR and relativistic quantum mechanics with
a common system of gauge principles. Besides facilitating the application
of quantum mechanics to astrophysics, this opens up new possibilities for
a grand unification of gravitation and electroweak theories, as explained
in Section V.
	
Claims about the felicity and power of GTG are amply supported by a steady
stream of recent publications. These include clarification and simplification
of Kerr solutions \cite{Doran02, Doran05, Lasenby05a, Dolan06}, new solutions
of the Dirac equation in a black
hole field \cite{Doran03b, Doran03c}, and new results from cosmological modeling
with conformal GA \cite{Lasenby04, Lasenby05b,Lasenby06a,Lasenby05c}. A few
words are in order to describe what these substantial publications have to offer.

In GTG the curvature tensor for a Kerr black hole with mass M and angular
momentum $L$ has the marvelously compact form:\cite{Hest05,Lasenby98}
\begin{equation}
R(B)=-\frac{M}{2(r-iL \cos\theta)^3}(B+3\hat{\bf x} B\hat{\bf x}),
\label{2.1}
\end{equation}
This expresses the curvature $R(B)$ as a bivector-valued function of a bivector
variable (explained below). As the notation suggests, $\hat{\bf x}$  is the
radial unit vector from the origin.

Of course, $R(B)$ reduces to the Schwarzschild curvature when $L = 0$. The
``complex" denominator is familiar from the literature, but only GTG reveals
that the unit imaginary $i$ is properly identified as a pseudoscalar. The
simplicity of $R(B)$ belies the subtlety of Kerr geometry and the unsolved
problem of matching it to a physically reasonable model of a rotating source.
New mathematical techniques to attack that problem for axisymmetric systems
are introduced in {\citenumber{Doran03b}}, where, among other things, asymptotic
uniqueness of the Kerr solution is proven.

A rigorous analysis of singularities in Kerr-Schild geometries is given in
{\citenumber{Doran03a}}.
Schwarzschild and Vaidya solutions (without maximal extensions!) are shown to
result from a delta-function point source. For the Reissner-Nordstrom solution
it is proved that gravity removes the divergent self-energy of classical
electromagnetism. Analysis of the Kerr solution at the source reveals a disk
of tension surrounded by a matter ring singularity, and physical significance
of this structure is discussed.

Turning to the important astrophysical problem of characterizing
fermion interactions with black holes: Its relativistic nature calls for an
attack using the Dirac equation. However, there is a paucity of such studies,
perhaps owing to clumsiness of the standard matrix methods, which have yielded
unsatisfactory results in the past. However, the STA version of the Dirac
equation is easier to solve than the matrix version, as demonstrated by
its use in the studies of black hole quantum mechanics
\cite{Doran02, Doran05, Lasenby05a, Dolan06}. Ironically,
 though each of the
calculations was first done with STA, some  calculations
had to be translated into matrix form to just satisfy journal referees.
In {\citenumber{Lasenby05a}}, for example, translation to matrix form took
longer than initial solution of the problem using STA.

The gravitational analog of the Mott formula for Coulomb scattering is derived
in {\citenumber{Doran02}} for the first time. This is extended to next order in
the Born series in {\citenumber{Dolan06}}, where numerical calculations
for absorption, scattering cross section and polarization are compared
with theoretical expectations. In {\citenumber{Doran05}}
absorption cross sections are calculated for a range of gravitational
couplings and initial conditions, and experimental signatures are studied in
detail. The gravitational analog of the hydrogen atom spectrum is derived
in {\citenumber{Lasenby05a}} for the first time. The eigenstates have complex energies.
 For small
couplings, real parts closely follow a hydrogen-like spectrum. Imaginary parts
give decay times due to absorption properties of the hole. Deformation of
the spectrum as coupling strength increases is thoroughly studied.

On the cosmological front, an extension of STA called conformal GA has been
used for a new approach to constructing inflationary models in closed
universes. Conformal embedding of closed-universe models in a de Sitter
background suggests a quantization condition on the available conformal
time that is explored in {\citenumber{Lasenby04,Lasenby05b}}. The result is a
predictive model producing
reasonable values for the major cosmological parameters and an excellent fit
to the most recent CMB radiation data. To fully appreciate the theoretical
depth of this work, one needs familiarity with the surprising subtleties of
conformal GA, which are nicely expounded in {\citenumber{Lasenby06a,Lasenby05c}}.

\section{Spacetime Algebra}\label{STA}

For physicists familiar with the Dirac matrix algebra, the
quickest approach to  STA is by reinterpreting the Dirac matrices as an {\it orthonormal
vector basis} $\{\gamma_\mu; \mu= 0,1,2,3\}$ for a 4D {\it real Minkowski vector space}
with signature specified by the rules:
\begin{equation}
\gamma_0^2=1 \qquad \hbox{and} \qquad \gamma_1^2=\gamma_2^2=\gamma_3^2=-1 \,.\label{3.1}
\end{equation}
The frame $\{\gamma_\mu\}$ generates the Spacetime Algebra (STA), an associative geometric algebra that is isomorphic to
the Dirac algebra (over the reals --- no complex scalars needed!). The product $\gamma_\mu\gamma_\nu$ of two vectors is called the {\it
geometric product.} The usual {\it inner product} of vectors is defined by
\begin{equation}
\gamma_\mu\cdot\gamma_\nu\equiv \frac{1}{2}(\gamma_\mu\gamma_\nu+\gamma_\nu\gamma_\mu)=
\eta_\mu\delta_{\mu\nu}  \,,\label{3.2}
\end{equation}
where $\eta_\mu=\gamma_\mu^2$ is the {\it signature indicator}. The {\it outer
product}
\begin{equation}
\gamma_\mu\wedge\gamma_\nu\equiv \frac{1}{2}(\gamma_\mu\gamma_\nu-\gamma_\nu\gamma_\mu)=
-\gamma_\nu\wedge\gamma_\mu  \,,\label{3.3}
\end{equation}
defines a new entity called a {\it bivector} (or 2-vector), which can be interpreted as a
directed plane segment representing the plane containing the two vectors.

 A complete basis for STA is given by the set:
\begin{tabbing}
\qquad $1$ \qquad\qquad \= $\{\gamma_\mu\}$ \qquad\qquad \=$\{\gamma_\mu\wedge\gamma_\nu\}$
\qquad\qquad
\=$\{\gamma_\mu i\}$ \qquad\qquad \=$i$ \\
\qquad 1 scalar \> 4 vectors \> 6 bivectors \> 4 trivectors \> 1 pseudoscalar\\
\qquad grade 0 \> grade 1 \> grade 2 \> grade 3 \> grade 4
\end{tabbing}
where the  {\it unit pseudoscalar}
\begin{equation}
i\equiv \gamma_0\gamma_1\gamma_2\gamma_3 \,\label{3.4}
\end{equation}
squares to $-1$, anticommutes with all odd grade elements and commutes with even grade
elements. Thus, STA is a linear space of dimension $1+4+6+4+1=2^4=16.$

A generic element in STA is called a {\it multivector}. Any multivector can be
expressed as a linear combination of the basis elements. For example, a
bivector $B$ has the expansion
\begin{equation}
B = \frac{1}{2} B^{\mu\nu}\gamma_\mu\wedge\gamma_\nu\,,\label{3.5}
\end{equation}
with its ``scalar components" $B^{\mu\nu}$ in the usual tensorial form. For components,
we use the usual tensor algebra conventions for raising and lowering indices and summing
over repeated upper and lower index pairs.

Any multivector $M$ can be written in the  {\it expanded form}
\begin{equation}
M = \alpha + a + B + bi + \beta{}i=\sum_{k=0}^4 \langle M\rangle_k,\label{3.6}
\end{equation}
where $\alpha=\langle M\rangle_0 \equiv \langle M\rangle$ and $\beta$ are scalars,
$a=\langle M\rangle_1$ and
$b$ are vectors, and $B=\langle M\rangle_2$ is a bivector, while $bi=\langle M\rangle_3$
is a trivector (or pseudovector) and $\beta{}i=\langle M\rangle_4$ is a pseudoscalar.

A multivector is said to be {\it even/odd} if it commutes/aniticommutes with the
unit pseudoscalar. From eqn. (\ref{3.6}) it is obvious that every multivector can be
expressed as a sum of even and odd parts.
The even multivectors compose a subalgebra of the STA
generated by the bivectors $\{\bsig_k \equiv \gamma_k\gamma_0; k= 1,2,3\}$, so that
\begin{equation}
\bsig_1\bsig_2\bsig_3= \gamma_0\gamma_1\gamma_2\gamma_3=i.\label{3.7}
\end{equation}

Coordinate-free computations are facilitated by various definitions, such as the operation
of {\it reversion}, which, for any multivector $M$ in the expanded form (\ref{3.6}),
can be defined by
\begin{equation}
\tilde{M} =\alpha + a - B - bi + \beta i\,.\label{3.8}
\end{equation}
This operation reverses the order in a product of vectors, so for vectors $a,b,c$
we have $(abc)\,\, \widetilde {} = cba$.

An even multivector $L$  subject to the normalization condition
\begin{equation}
L\tilde{L}  = 1  \label{3.9}
\end{equation}
is called a {\it rotor}. It follows that every rotor can be expressed in the exponential
form
\begin{equation}
L= e^{\frac{1}{2}B}\,,\label{3.10}
\end{equation}
where $B$ is a bivector.
The rotors form a multiplicative group variously called the  {\it spacetime
rotor group} or {\it spin group} Spin$_+$(1,3), SU(2C),
or the ``spin representation" of the Lorentz rotation group.

The concepts of vector and multivector have been defined by algebraic rules specified
above. Alternatively, vectors are often defined by their behavior under a transformation
group. To keep these two concepts separate, we will use the term {\it tensor} to refer
to the latter. More generally, we distinguish two kinds of multivectors, {\it tensors}
and {\it spinors}, by the way they transform.
Under the rotor Lorentz group $\{L\}$, a {\it tensor F} obeys the transformation law
\begin{equation}
L:   F\quad\rightarrow\quad  F'= LF\tilde{L}\,,\label{3.11}
\end{equation}
while a {\it spinor} $\psi$ obeys
\begin{equation}
L:  \psi\quad\rightarrow\quad  \psi'= L\psi \,.\label{3.12}
\end{equation}
Evidently spinors are more fundamental than tensors, because every spinor $\psi$ determines
the tensor transformation law
\begin{equation}
L:  \psi\Gamma\tilde{\psi}\quad\rightarrow\quad  \psi'\Gamma\tilde{\psi'}
= L(\psi\Gamma\tilde{\psi})\tilde{L} \,\label{3.13}
\end{equation}
for any fixed multivector $\Gamma$, while one cannot construct spinors from tensors.

\section{Real Quantum Mechanics }\label{RQM}

In real QM, the {\it electron wave function} $\psi=\psi(x)$ is a spinor field
in the real STA. For the electron with charge $e$ and mass $m_e$,
the field equation for $\psi$ is the {\it real Dirac equation}\cite{Hest03b}
\begin{equation}
\gamma^\mu\partial_\mu(\psi i\bsig_3\hbar-eA_\mu\psi)=m_e\psi\gamma_0\,,\label{4.1}
\end{equation}
or, in manifestly coordinate-free form,
\begin{equation}
\partial\psi i\bsig_3\hbar-eA\psi=m_e\psi\gamma_0\,,\label{4.2}
\end{equation}
where $A=A_\mu\gamma^\mu$ is the electromagnetic vector potential and the {\it vector
derivative} $\partial \equiv\gamma^\mu\partial_\mu$ will be recognized as the
famous differential operator introduced by Dirac, except that the $\gamma^\mu$ are
vectors rather than matrices.
This equation is isomorphic to the standard matrix version of the Dirac equation, so it
has identical physical content. However, it has the great advantage of representing geometric
structure explicitly that is inherent but hidden in the matrix version.
In particular, the unit imaginary in the matrix version is explicitly identified
 with the spacelike bivector $i\bsig_3=\gamma_2\gamma_1$, which indeed does square to $-1$.

The {\it real spinor field} $\psi=\psi(x)$ in (\ref{4.1}) is an even multivector with
the invariant form\cite{Hest03b}
\begin{equation}
\psi =(\rho e^{i\beta})^{1/2}R,\label{4.3}
\end{equation}
where $R=R(x)$ is a rotor field and $\rho$ and $\beta$ are scalar-valued functions. From (\ref{3.10}) it is clear
that $R$ is a 6-parameter function, so $\psi$ has eight degrees of freedom, as required for
equivalence with the standard matrix form of the Dirac wave function. This kind of spinor is
called an {\it operator spinor}, because it operates on fixed vectors to create tensor
``observables."

Thus, the Dirac wave function determines a set of {\it local observables}
\begin{equation}
\psi\gamma_\mu\tilde{\psi}= \rho e_\mu \,,\label{4.4}
\end{equation}
where $\rho=\rho(x)$ is a scalar probability density and
\begin{equation}
 e_\mu =R\gamma_\mu \tilde{R} \,,\label{4.5}
\end{equation}
is a frame field of orthonormal vectors. The vector field $\psi\gamma_0\tilde{\psi}= \rho e_0$ is the
{\it Dirac probability current}, which doubles as a charge current when multiplied by the
charge $e$. The vector field $e_3 =R\gamma_3 \tilde{R}$ specifies the local direction of
electron spin. The vector fields $e_1$ and $e_2$ specify the local phase of the
electron, and $e_2e_1= R\gamma_2\gamma_1\tilde{R}$ relates the unit imaginary in the Dirac
equation to electron spin. Full justification for these interpretations is given elsewhere.
\cite{Hest03c,Hest03b}

We need several other features of standard QM translated into real Dirac theory.
Multiplication of the Dirac equation (\ref{4.1}) on the right by $\bsig_2$ has a net effect of
changing the sign of the electric charge. So (suppressing an inconsequential phase factor)
we can define a {\it positron wave function} $\psi^C\equiv\psi\bsig_2$ that satisfies
the {\it charge conjugate} equation
\begin{equation}
\gamma^\mu(\partial_\mu\psi^C i\bsig_3\hbar+eA_\mu\psi^C)=m_e\psi^C\gamma_0\,.\label{4.6}
\end{equation}

In the matrix theory a {\it chirality operator} is defined by
$\gamma_5\equiv -i'\gamma_0\gamma_1\gamma_2\gamma_3$ where $i'$ is the unit imaginary in
the matrix theory. Recalling the definition of the unit pseudoscalar $i$, we translate
this into a real operator defined by
\begin{equation}
\gamma_5\psi\equiv i\psi(-i\bsig_3)=\psi\bsig_3.\label{4.7}
\end{equation}
This enables a chiral decomposition of the wave function
\begin{equation}
\psi=\psi_l+\psi_r,\label{4.8}
\end{equation}
where left-handed and right-handed chiral eigenstates are defined, respectively by
\begin{equation}
\psi_l\equiv \psi\frac{1}{2}(1-\bsig_3)=-\psi_l\bsig_3, \label{4.9}
\end{equation}
\begin{equation}
\psi_r\equiv \psi\frac{1}{2}(1+\bsig_3)=\psi_r\bsig_3, \label{4.10}
\end{equation}
The usual Lagrangian for a Dirac electron can be put in the form
\begin{eqnarray}
\mathcal{L}_e&=&<(\partial\psi i\bsig_3\hbar-eA\psi)\gamma_0\tilde{\psi}-m\psi\tilde{\psi}>
=\hbar<-\partial\psi_l i\gamma_0\tilde{\psi_l}+\partial\psi_r i\gamma_0\tilde{\psi_r}>
\nonumber\\
& &-<eA(\psi_l\gamma_0\tilde{\psi_l}+\psi_r\gamma_0\tilde{\psi_r})
+m_e(\psi_l\tilde{\psi_r}+\psi_r\tilde{\psi_l})>
\,,\label{4.11}
\end{eqnarray}
where the angular brackets signify ``scalar part." Note that left and right-handed components
are coupled only in the mass term. We shall see that this relates to the Higgs mechanism for
generating mass.

The electron Lagrangian is invariant under the electromagnetic gauge transformation:
\begin{eqnarray}
\psi\quad&\rightarrow&\quad  \psi'= \psi e^{i\bsig_3 \chi}\,.\nonumber\\
eA\quad&\rightarrow&\quad  eA'= eA-\partial \chi\,.\label{4.12}
\end{eqnarray}
 This shows explicitly that the bivector $i\bsig_3 =\gamma_2\gamma_1$ is the generator
of electromagnetic gauge transformations in the Dirac equation. It follows that gauge
transformations are expressions of spacetime geometry in real QM.
We shall see that this insight has implications for electroweak gauge theory.

\section{Gauge Theory Gravity with Real Quantum Mechanics}\label{GTG}

In {\it Gauge Theory Gravity} (GTG) developed by Lasenby, Doran and Gull,\cite{Lasenby98}
the fundamental geometric entity is the Dirac spinor rather than the line element of GR.
This greatly simplifies the integration of gravity theory with quantum mechanics.
The Dirac equation (\ref{4.1}) generalizes immediately to
\begin{equation}
g^\mu(D_\mu\psi i\bsig_3\hbar-eA_\mu\psi)=m_e\psi\gamma_0\,,\label{5.1}
\end{equation}
where a {\it coordinate frame} of vector fields is defined by
\begin{equation}
g^\mu=h_a^\mu\gamma^a,\label{5.2}
\end{equation}
a spinor {\it coderivative} $D_\mu$ is defined by
\begin{equation}
D_\mu\psi= (\partial_\mu +\frac{1}{2}\omega_\mu)\psi,\label{5.3}
\end{equation}
with a bivector-valued {\it ``spin connexion"} defined by
\begin{equation}
\omega_\mu=\frac{1}{2}\omega_\mu^{ab}\gamma_a\wedge\gamma_b.\label{5.4}
\end{equation}
This will be recognized as formally identical to the usual formulation of the Dirac
equation in GR, with the coefficients $h_a^\mu=g^\mu\cdot\gamma_a$ identified as
components of a {\it vierbein}. So what's new besides the fact that $\{\gamma^a\}$
is a frame of constant vectors instead of matrices?
Only a brief outline of the answer can be given here; full details are given in
the references to GTG already cited.

GTG formulates gravity as a gauge field theory on the flat background of Minkowski
spacetime. Each spactime point $x$ is a vector that can be parametrized
locally by an arbitrary set of coordinates $\{x^\mu\}$. Without loss of generality we
can employ rectangular coordinates so that $x=x^\mu\gamma_\mu$ and
$\partial_\mu x=\gamma_\mu$.

In GTG Einstein's principles of covariance and equivalence are replaced, respectively,
by two gauge principles:
\smallskip

\noindent {\bf{}Displacement Gauge Invariance:}
 {\it The equations of physics must be invariant under arbitrary smooth remappings
of events onto spacetime.}
\smallskip

\noindent{\bf{}Rotation Gauge Covariance:}
 {\it The equations of physics must be covariant under local Lorentz rotations.}
\smallskip

\noindent These principles have well-defined mathematical implementations.
Displacement invariance requires existence of a differentiable {\it gauge tensor}
$\bar{h}$ so that
\begin{equation}
g^\mu=\bar{h}(\gamma^\mu)=h_a^\mu\gamma^a,\label{5.5}
\end{equation}
This elevates the vierbein from an auxiliary quantity to a fundamental theoretical entity.

The gauge tensor must be nonsingular, so it has an inverse
\begin{equation}
g_\mu=\underline{h}^{-1}(\gamma_\mu).\label{5.5b}
\end{equation}
The usual {\it metric tensor} is determined by the gauge tensor:
\begin{equation}
g_{\mu\nu}\equiv g_\mu\cdot g_\nu=\underline{h}^{-1}(\gamma_\mu)\cdot
\underline{h}^{-1}(\gamma_\nu).\label{5.6}
\end{equation}
The metric tensor is now an auxiliary quantity.

Rotation gauge covariance is implemented by requiring local covariance of tensors and
spinors under the Lorentz rotations (\ref{3.11}) and (\ref{3.12}) where $L=L(x)$ is an
arbitrary position dependent rotor field.

The usual gauge covariant derivative (\ref{5.3}) is then defined by by introducing
the connexion (\ref{5.4}) so that
\begin{equation}
L:  D_\mu\psi\quad\rightarrow\quad  L(D_\mu\psi)=
D'_\mu\psi'=(\partial_\mu+\frac{1}{2}\omega'_\mu)\psi'\,.\label{5.7}
\end{equation}
Accordingly, the connexion must obey the familiar gauge theory transformation law
\begin{equation}
L: \omega_\mu=\omega (g_\mu)\quad\rightarrow\quad
\omega'_\mu=L\omega_\mu\tilde{L}-2(\partial_\mu L)\tilde{L}\,.\label{5.8}
\end{equation}
As the gauge field $g^\nu=\bar{h}(\gamma^\nu)$ is a tensor field, its
coderivative is given by
\begin{equation}
D_\mu g^\nu=\partial_\mu g^\nu+\omega_\mu\times g^\nu\,,\label{5.8b}
\end{equation}
where the {\it commutator product} is defined by $A\times B\equiv\frac{1}{2} (AB-BA)$.

The {\it curvature tensor}, expressed as a bivector-valued linear function of a
bivector variable $R(a\wedge b)=a^\mu b^\nu R(g_\mu\wedge g_\nu)$,
is obtained immediately from the commutator of coderivatives. Thus,
\begin{equation}
[D_\mu, D_\nu] \psi=\frac{1}{2} R(g_\mu\wedge g_\nu)\psi\,,\label{5.9}
\end{equation}
where
\begin{equation}
R(g_\mu\wedge g_\nu)=\partial_\mu\omega_\nu-\partial_\nu\omega_\mu
+\omega_\mu\times\omega_\nu. \label{5.10}
\end{equation}
This is the fundamental form for curvature in GTG.
It follows that the curvature bivector has the gauge covariance property
\begin{equation}
L:  R(a \wedge b)\quad\rightarrow\quad  R'(a' \wedge b') =
L R(L(a \wedge b)\tilde{L})\tilde{L}.\label{5.11}
\end{equation}

Gravitational field equations can be obtained from a Lagrangian.\cite{Lasenby98}
 As GTG postulates
two basic variables, the gauge field and the connexion, there must be two field equations.
The first is equivalent to Cartan's ``first fundamental form:"
\begin{equation}
D\wedge g^\nu\equiv g^\mu\wedge D_\mu g^\nu=\theta^\nu,\label{5.12}
\end{equation}
where the {\it torsion}  $\theta^\nu=\theta(g^\nu)$ is a linear bivector-valued
function of a vector field. In GR the torsion vanishes, in which case the connexion
is Riemannian, and (\ref{5.12}) can be solved algebraically to give the connexion
as a function of the gauge tensor and its first derivatives. However, it has been shown
\cite{Lasenby98} that a matter field described by the Dirac equation gives rise to
a non-vanishing torsion.

The other field equation is, of course, {\it Einstein's equation}:
\begin{equation}
G(g^\mu) = \kappa T(g^\mu),\label{5.13}
\end{equation}
where components of the matter energy-momentum tensor are given by
$T^{\nu\mu}=g^\nu\cdot T(g^\nu)$.
For {\it Einstein's tensor}, STA gives us the elegant compact form
\begin{equation}
G(g^\mu)=\frac{1}{2}(g^\mu \wedge g^\alpha \wedge g^\beta)
\cdot R(g_\alpha \wedge g_\beta).\label{5.14}
\end{equation}

\section{Gravelectroweak Gauge Theory: Standard Model}\label{Gew}

The {\it gravelectric Dirac equation} (\ref{5.1}) is covariant under the {\it gravelectric}
gauge transformation:
\begin{equation}
\psi\quad\rightarrow\quad  \psi'= L\psi U,\label{6.1}
\end{equation}
where $L$ is a Lorentz rotor and $U= e^{i\bsig_3 \chi}$. This can be regarded as a spacetime
transformation acting on the left of $\psi$ and an (internal) {\it isospace} transformation
acting on the right.

According to standard electroweak theory, the electromagnetic gauge group is only
one part of the electroweak group. But real Dirac theory tells us that the
 electromagnetic gauge generator is a geometric object $i\bsig_3$ acting on the right
of $\psi$,
so we should expect similar geometric objects generating the rest of the electroweak group.
Indeed, the most general gauge transformation covariant on the right side of (\ref{5.1})
satisfies
\begin{equation}
\tilde{U}\gamma_0 U = \gamma_0 \label{6.2}
\end{equation}
The general solution of this equation has the form
\begin{equation}
U=e^{\frac{1}{2}i\,\btheta} e^{\frac{1}{2}i\,\chi}, \label{6.3}
\end{equation}
where $\btheta=\theta_1 \bsig_1+\theta_2 \bsig_2+\theta_3 \bsig_3$. This is exactly the
gauge group SU(2)$\otimes$U(1) of electroweak theory. It strongly suggests that the
geometric structure of electroweak theory is already inherent in Dirac theory!

Thus we arrive at the question: How can we revise the Dirac equation (\ref{5.1}) to
incorporate electroweak interactions acting on the right side of the wave function?
 We shall consider two answers.
In this section we see how the structure of the standard model can be incorporated
in a way that links it to spacetime geometry with new physical features.
In a later section we formulate a simpler alternative to the standard model that certainly
will have new physical predictions.

According to the standard model electroweak interactions couple differently to the
two chiral eigenstates of the electron. For that purpose, a slick way to separate chiral
eigenstates is to
multiply the electron wave function $\psi$ by the {\it idempotent} (a projection operator)
$\frac{1}{2}(1+\gamma_0)(1-\bsig_3)$ to get
\begin{equation}
\Psi=\psi\frac{1}{2}(1+\gamma_0)(1-\bsig_3)=\psi_l+\psi_r\gamma_0.\label{6.4}
\end{equation}
This type of spinor is called an {\it ideal spinor}, because its values are elements of a {\it
minimal left ideal} in the STA.
Note that the chiral components of the operator spinor (\ref{4.8}) have been neatly separated
into even and odd parts of the ideal spinor (\ref{6.4}).

Multiplying (\ref{5.1}) by the same idempotent, we get the Dirac equation in the ``ideal
form"
\begin{equation}
-g^\mu(D_\mu\Psi i\hbar+eA_\mu\Psi)=m_e\Psi.\label{6.5}
\end{equation}
This looks more like the matrix Dirac equation, though with the pseudoscalar $i$
 playing the role of unit imaginary. It is, in fact, the first version of the Dirac
equation forumlated in the real STA.\cite{Hest66} It was noted at the time
that minimal ideals can be used to incorporate isospace degrees of freedom. However,
physical motivation for implementing this possibility had to wait for development
of electroweak theory.
The key insight was noticing the separation of chiral eigenstates in (\ref{6.4}).
That paved the way for a leptonic version of Dirac's equation that incorporates
all the electroweak interactions.\cite{Hest82}

For the purpose of electroweak theory, we affix a subscript to the electron wave function
\begin{equation}
\Psi_e=\psi_e\frac{1}{2}(1+\gamma_0)(1-\bsig_3)=\psi_l+\psi_r\gamma_0
=-\Psi_e \bsig_3,\label{6.6}
\end{equation}
and we define a neutrino wave function\cite{Hest82,Doran03}
\begin{equation}
\Psi_\nu=\psi_\nu\frac{1}{2}(1-\bsig_3)\bsig_1=\psi_\nu\frac{1}{2}\bsig_1(1+\bsig_3)
=\Psi_\nu \bsig_3.\label{6.7}
\end{equation}
The right side of these equations shows that the $\Psi_e$ and $\Psi_\nu$ are different
 eigenstates of $\bsig_3$ operating from the right. In other words, they are elements of
independent minimal left ideals.
The neutrino wave function in (\ref{6.7}) is left-handed, with no right-handed (or odd)
component. In common parlance, it represents a 2-component Weyl neutrino.

Now we regard electron and neutrino as different states of a single lepton with wave function
\begin{equation}
\Psi=\Psi_e+\Psi_\nu=\Psi_l+\Psi_r.\label{6.8}
\end{equation}
The left-handed and right-handed components are defined, respectively, by
\begin{equation}
\Psi_l\equiv\frac{1}{2}(\Psi-i\Psi i)= \psi_e\frac{1}{2}(1-\bsig_3)
+\psi_\nu \frac{1}{2}(1-\bsig_3)\bsig_1, \label{6.9}
\end{equation}
\begin{equation}
\Psi_r\equiv\frac{1}{2}(\Psi+i\Psi i)= \psi_e\frac{1}{2}(1+\bsig_3)\gamma_0. \label{6.10}
\end{equation}

Following the standard model, we assume a {\it gravelectroweak gauge transformation}
of the form
\begin{equation}
\Psi\quad\rightarrow\quad  \Psi'= L(\Psi_l U+\Psi_r \tilde{U} U),\label{6.11}
\end{equation}
where $U$ is defined as before in (\ref{6.3}), so  $\tilde{U} U=e^{i\,\chi}$. One says that
$\Psi_l$ transforms as an {\it isospinor} (on the right), while $\Psi_r$ transforms
as an {\it isoscalar}.

Now we can define a generalized coderivative in the usual way:
\begin{equation}
\mathcal{D}_\mu\Psi=D_\mu\Psi-\Psi_l iW_\mu-\Psi_r ig'B_\mu,\label{6.12}
\end{equation}
with $\hbar=1$ from here on.
In perfect analogy to the gravitational connexion (\ref{5.8}) the nonabelian electroweak
connexion obeys the transformation law
\begin{equation}
W_\mu\quad\rightarrow\quad  W_\mu'= \tilde{U}W_\mu U-\tilde{U} \partial_\mu U.\label{6.11}
\end{equation}
Using notation of the standard model as closely as possible, the connexion
has the form
\begin{eqnarray}
2W_\mu&\equiv& g\mathbf{A}_\mu-g'B_\mu\nonumber\\
&=&\frac{g}{\sqrt{2}}[W^+_\mu(1+\bsig_3)\bsig_1+W^-_\mu(1-\bsig_3)\bsig_1]
+gA^3_\mu\bsig_3-g'B_\mu,\label{6.13}
\end{eqnarray}
where
\begin{equation}
W^\pm_\mu\equiv \frac{1}{\sqrt{2}}(A^1_\mu\mp iA^2_\mu).\label{6.13b}
\end{equation}
Introducing the {\it Weinberg angle} in the standard way, we have
\begin{equation}
gA^3_\mu\bsig_3-g'B_\mu=-eA_\mu(1-\bsig_3)-\frac{g}{\cos\theta_W}
Z_\mu(\bsig_3\cos^2\theta_W+\sin^2\theta_W),\label{6.14}
\end{equation}
\begin{equation}
g'B_\mu=eA_\mu+\frac{g}{\cos\theta_W}Z_\mu\sin^2\theta_W,\label{6.15}
\end{equation}
where $e=g\sin\theta_W=g'\cos\theta_W$ relates the electron charge to the weak
coupling constants $g$ and $g'$.

It is convenient to put all this together in a single lepton Lagrangian:
\begin{eqnarray}
h\mathcal{L}_{lep}&=&<\mathcal{D}\Psi i\gamma_0\tilde{\Psi}>
\equiv<g^\mu\mathcal{D}_\mu\Psi i\gamma_0\tilde{\Psi}>\nonumber\\
&=&<D\Psi i\gamma_0\tilde{\Psi}>+<g^\mu(\Psi_l W_\mu+\Psi_r g'B_\mu)\gamma_0\tilde{\Psi}>\nonumber\\
&=&<D\Psi i\gamma_0\tilde{\Psi}>+\frac{g}{2\sqrt{2}}(W^+J_-+W^-J_+)
-eA\cdot J+\frac{g}{\sqrt{2}}Z\cdot J_Z
.\label{6.16}
\end{eqnarray}
The sundry quantities in this Lagrangian are defined in the following. The scale factor
$h\equiv \det \underline{h}$ is necessary to make $\mathcal{L}_{lep}$ a gauge
invariant Langrangian density.

The kinetic-gravitation component of the Lagrangian has the familiar form
\begin{equation}
<D\Psi i\gamma_0\tilde{\Psi}>=<g^\mu (\partial_\mu
+\frac{1}{2}\omega_\mu)\Psi i\gamma_0 \tilde{\Psi}>,\label{6.17}
\end{equation}
which we will not explicate further, as we are most interested in the electroweak
interactions.

As usual, the electromagnetic vector potential is $A=g^\mu A_\mu$. The weak gauge fields
are the charged fields $W^+=g^\mu W^+_\mu=\tilde{W}^-$ (having vector and pseudovector parts)
and the neutral vector field $Z=g^\mu Z_\mu$.
After some fairly easy computations (exploiting the fact that odd terms
cannot have a scalar part), we arrive at the following explicit forms for the
corresponding {\it lepton currents}.
The usual Dirac current is given by
\begin{equation}
J=\psi_e\gamma_0\tilde{\psi}_e= \psi_e\frac{1}{2}(\gamma_0 -\gamma_3)\tilde{\psi}_e
+\psi_e\frac{1}{2}(\gamma_0 +\gamma_3)\tilde{\psi}_e, \label{6.18}
\end{equation}
where the right side separates the contributions of left-and right-handed components.
The charged and neutral weak currents are
\begin{equation}
J_-= \psi_e(\gamma_0 -\gamma_3)\tilde{\psi}_\nu = \tilde{J}_+, \label{6.19}
\end{equation}
\begin{equation}
J_Z= \psi_\nu(\gamma_0 -\gamma_3)\tilde{\psi}_\nu
-\psi_e\frac{1}{2}(\gamma_0 -\gamma_3- 4 \sin^2 \theta_W \gamma_0)\tilde{\psi}_e. \label{6.20}
\end{equation}
This completes our characterization of the {\it gravelectroweak Lagrangian} (\ref{6.16})
for leptons.

The rest of the standard model is now easily transcribed into STA. Lagrangians for
the gauge fields carry over with no essential change. The Higgs mechanism has
been transcribed by Antony Lewis in unpublished notes.\cite{Lewis98}

The electron mass can be incorporated in the Lagrangian (\ref{6.16}) by subtracting the term
\begin{equation}
<\Psi\underline{m}\tilde{\Psi}> = m_e<\psi_e \tilde{\psi}_e>, \label{6.21}
\end{equation}
Higgs theory replaces the mass operator $\underline{m}=m_e\gamma_0$ by a Higgs field.
As shown explicitly in (\ref{4.11}), the mass (hence the Higgs mechanism) requires
coupling between left- and right-handed electron states. Therefore, if a small neutrino mass
is to arise from the same mechanism, a right-handed component must be added to the neutrino
wave function (\ref{6.7}).
In that case, the mass operator would generalize to
\begin{equation}
\underline{m}=[m_e\frac{1}{2}(1 -\bsig_3)+m_\nu\frac{1}{2}(1 +\bsig_3)]\gamma_0, \label{6.21b}
\end{equation}
 where $m_\nu$ is the neutrino mass.

Looking back, we see that Dirac theory does indeed have degrees of freedom
sufficient to incorporate electroweak theory in a fairly straightforward way.
{\bf One thing new stands out:} the idempotent $\frac{1}{2}(1 -\bsig_3)$ that breaks the
symmetry of the electroweak connexion in (\ref{6.13}) by separating electromagnetic from weak
interactions, also separates left and right-handed lepton states in (\ref{6.9}) and
(\ref{6.10}). In the standard model this relation between symmetry breaking and
chiral eigenstates is incorporated as an independent assumption.
In this respect, therefore, electroweak theory is simplified by its geometrization in STA.
Though this may not produce new physical predictions, it does reveal new
possibilities for modifying electroweak theory that surely will produce new predictions.
One such possibility is developed next.

\section{Zitterbewegung in Dirac Theory}\label{ZDT}

Having aligned the standard model with geometry of the Dirac equation, we notice that one
prominent feature of Dirac theory is missing, namely, the {\it zitterbewegung} (zbw) of the
electron. The zbw was discovered and given its name by Schroedinger in an analysis of
free particle solutions of the Dirac equation.\cite{Schr30}
It has since been recognized as a general feature of electron phase fluctuations and proposed
as a fundamental principle of QM.\cite{Hest90,Hest03c}

One reason that the significance of zbw has been consistently overlooked, especially in
electroweak theory, is that the relevant observables are not among the so-called ``bilinear
covariants," from which observable currents are constructed in the standard model. I refer
to the vector fields $\psi\gamma_1\tilde{\psi}= \rho e_1$ and
$\psi\gamma_2\tilde{\psi}= \rho e_2$, identified as observables in (\ref{4.4}).
The standard model deals only with $\psi\gamma_0\tilde{\psi}= \rho e_0$
 and $\psi\gamma_3\tilde{\psi}= \rho e_3$, especially in combination to form
chiral currents in
(\ref{6.18}), (\ref{6.19}), and (\ref{6.20}).

One way to recognize the significance of the observables $ e_2$ and  $e_1$ is to note that
they rotate with twice the electron phase along streamlines of the conserved Dirac current.
The rotation rate for a free electron is the  $2m_e/\hbar\approx 1$ ZHz (zettaHertz), the
{\it zbw frequency} found by Schroedinger. This rate varies in the presence of interactions
 but still remains outside the range of direct observation.

Other features of the zbw may be detectible, however. In particular, it has often been
 suggested that the electron's magnetic moment is generated
by a circulating charged current. That suggestion is elevated to a principle by replacing
the charged Dirac current $e\psi\gamma_0\tilde{\psi}$ by the {\it zbw current}
$e\psi(\gamma_0-\gamma_2)\tilde{\psi}= e\rho (e_0-e_2)$.
Obviously, the zbw current is analogous to the left-handed chiral current
$\psi_e\frac{1}{2}(\gamma_0 -\gamma_3)\tilde{\psi}_e$ with $\gamma_3$ replaced by $\gamma_2$.
Equation (\ref{6.18}) expresses the Dirac current as a sum of left- and right-handed chiral
currents. Therefore, we can incorporate zbw into electroweak theory by dropping the
right-handed current and replacing the left-handed current  by the zbw current.

There is no need for a zbw analog of the right-handed chiral current.
Looking over the standard model in the preceding section, it is evident that the
right-handed current plays only a minor role. Its main function is to balance the
left-handed current to produce the Dirac current, as shown in (\ref{6.18}).
The theory can be simplified considerably once that function is seen to be unnecessary.
We see how in the next section.

\section{Gravelectroweak Gauge Theory: Majorana Model}\label{MM}

In Section \ref{Gew} we saw how the standard model can be incorporated into real Dirac
theory using the {\it chiral projection operator} $\frac{1}{2}(1-\bsig_3)$.
By a completely analogous procedure we now construct an alternative model using the
{\it Majorana projection operator} $\frac{1}{2}(1-\bsig_2)$, so named for reasons
that are obvious below.

As in Section \ref{RQM}, we begin with a real Dirac wave function $\Psi$, but here we
identify it as a lepton wave function with electron and neutrino
components defined as follows:
\begin{equation}
\Psi= \Psi_e+\Psi_\nu=\psi_e\frac{1}{2}(1-\bsig_2)
+\psi_\nu \frac{1}{2}(1-\bsig_2)\bsig_3,\label{8.1}
\end{equation}
with
\begin{equation}
\Psi_e=\frac{1}{2}(\Psi-\Psi\bsig_2)=\psi_e\frac{1}{2}(1-\bsig_2)=-\Psi_e \bsig_2
,\label{8.2}
\end{equation}
\begin{equation}
\Psi_\nu=\frac{1}{2}(\Psi+\Psi\bsig_2)=
\psi_\nu \frac{1}{2}(1-\bsig_2)\bsig_3=\Psi_\nu\bsig_2.\label{8.3}
\end{equation}
Note the analogy with equations (\ref{6.6}) and (\ref{6.7}). Here $\Psi_e$ and $\Psi_\nu$
are eigenstates of $\bsig_2$ operating on the right, which, as is evident from (\ref{4.6}),
makes them eigenfunctions of the charge conjugation operator, commonly called
{\it Majorana states} in the literature.

Note that we have introduced a major shift in physical interpretation of the Dirac
wave function, decomposing it into charged and neutral components
rather than the usual decomposition into positive and negative energy components.
This shift is not so radical as it might seem at first, because
it is completely analogous to the decomposition of (\ref{6.9}) in the standard model.
Of course, we have to modify interactions in the Dirac equation to be consistent with
the new interpretation, but those modifications are the same as in the standard model.

With the change in physical interpretation of the Dirac wave function comes a change
in physical interpretation of operators.
In particular, the {\it antiparticle conjugate} of the lepton wave function is
\begin{equation}
\Phi \equiv \Psi \bsig_3 =\varphi_{\bar{e}} \frac{1}{2}(1-\bsig_2)
+\varphi_{\bar{\nu}} \frac{1}{2}(1-\bsig_2)\bsig_3.\label{8.4}
\end{equation}

As in the standard model, we assume a {\it gravelectroweak gauge transformation}
of the form
\begin{equation}
\Psi\quad\rightarrow\quad  \Psi'= L\Psi U.\label{8.5}
\end{equation}
As before, the generalized coderivative for this gauge group has the form
\begin{equation}
\mathcal{D}_\mu\Psi=D_\mu\Psi-\Psi iW_\mu,\label{8.6}
\end{equation}
where $D_\mu=\partial_\mu+\frac{1}{2}\omega_\mu$ and the electroweak connexion
is given by
\begin{eqnarray}
2W_\mu&\equiv& g\mathbf{A}_\mu-g'B_\mu
=\frac{g}{\sqrt{2}}[W^+_\mu(1+\bsig_2)\bsig_3+W^-_\mu(1-\bsig_2)\bsig_3]\nonumber\\
& &-eA_\mu \frac{1}{2}(1-\bsig_2)
-\frac{g}{\cos\theta_W}Z_\mu(\bsig_2\cos^2\theta_W+\sin^2\theta_W),\label{8.7}
\end{eqnarray}
where
\begin{equation}
W^\pm_\mu\equiv \frac{1}{\sqrt{2}}(A^3_\mu\mp iA^1_\mu).\label{8.8}
\end{equation}
This differs from the standard model only in a permutation of indices due to splitting
the connexion with a Majorana projection instead of a chiral projection. All variables
and parameters are the same as in the standard model.

A modified Dirac equation for the Majorana model can now be written down immediately:
\begin{equation}
g^\mu\mathcal{D}_\mu\Psi i\gamma_3=\Psi \underline{m}\,,\label{8.9}
\end{equation}
where lepton masses have been incorporated in a mass operator
\begin{equation}
\underline{m}=m_e \frac{1}{2}(1-\bsig_2)
+m_\nu \frac{1}{2}(1+\bsig_2).\label{8.10}
\end{equation}
Of course, the mass operator can be replaced by a Higgs mechanism, but we leave that
as an open question. There are good reasons to believe that gravity plays a role in
determining particle masses.\cite{Ros05}
Unified gravelectroweak theory should help in investigating that possibility.

In constructing a Lagrangian to give us the field equation (\ref{8.9}), we note
a problem: The mass term in the usual Dirac Lagrangian (\ref{4.11}) requires a
4-component wave function, whereas the electron wave function (\ref{8.2}) here has
only 2 components.  A solution is suggested by noting that the usual mass term in
(\ref{4.11}) couples positive and negative energy states, so the analog here is coupling
lepton and antilepton states. Accordingly, we use the antilepton wave function (\ref{8.4})
to construct a Lagrangian of the form
\begin{equation}
h\mathcal{L}_{Maj}=<g^\mu\mathcal{D}_\mu\Psi i\gamma_3\tilde{\Phi}>
-<\Psi \underline{m}\tilde{\Phi}>\,,\label{8.11}
\end{equation}
Alternatively, we can use $\tilde{\Phi}=-\bsig_3\tilde{\Psi}$ to write the Lagrangian as
\begin{eqnarray}
h\mathcal{L}_{Maj}&=&<g^\mu\mathcal{D}_\mu\Psi i\gamma_0\tilde{\Psi}>
+<\Psi \underline{m}\bsig_3\tilde{\Psi}>\,\nonumber\\
&=&<D\Psi i\gamma_0\tilde{\Psi}>+<g^\mu\Psi W_\mu\gamma_0\tilde{\Psi}>
+<\Psi \underline{m}\bsig_3\tilde{\Psi}>.\label{8.12}
\end{eqnarray}
The last line singles out the electroweak interactions for further discussion.
It is interesting to note that
\begin{equation}
2W_\mu\gamma_0= g\mathbf{A}_\mu\gamma_0-g'B_\mu\gamma_0
= gA^k_\mu\gamma_k-g'B_\mu\gamma_0\label{8.13}
\end{equation}
is a spacetime vector, with 4 gauge fields coresponding to 3 spacelike components and a
distinguished timelike component.

The electroweak term in (\ref{8.12}) has exactly the same form as in the standard model
(\ref{6.16}), namely,
\begin{equation}
<g^\mu\Psi W_\mu\gamma_0\tilde{\Psi}>=\frac{g}{2\sqrt{2}}(W^+J_-+W^-J_+)
-eA\cdot J+\frac{g}{\sqrt{2}}Z\cdot J_Z.\label{8.14}
\end{equation}
The gauge fields $W^\pm, A, Z$ are the same as before, but the lepton currents are
somewhat different:
The  Dirac current is replaced  by the {\it zbw current}
\begin{equation}
J= \psi_e\frac{1}{2}(\gamma_0 -\gamma_2)\tilde{\psi}_e. \label{8.15}
\end{equation}
The charged and neutral weak currents are
\begin{equation}
J_-= \psi_e(\gamma_0 -\gamma_2)\tilde{\psi}_\nu = \tilde{J}_+, \label{8.16}
\end{equation}
\begin{equation}
J_Z= \psi_\nu(\gamma_0 -\gamma_2)\tilde{\psi}_\nu
-\psi_e\frac{1}{2}(\gamma_0 -\gamma_2- 4 \sin^2 \theta_W \gamma_0)\tilde{\psi}_e. \label{8.17}
\end{equation}
This completes our characterization of the {\it Majorana Lagrangian} (\ref{8.12})
for leptonic gravelectroweak interactions.

So far we have ignored the derivation of field equations from the Lagrangian
density, but that involves a subtle point that is often overlooked. Though the point
has been made before,\cite{Lasenby98} it is worth reviewing to complete the present theory.
 Regarding $\Psi$ and $\tilde{\Psi}$ as independent variables in the Lagrangian
(\ref{8.12}) appears to give the field equation (\ref{8.9}) immediately when
$\tilde{\Psi}$ is varied. But the correct procedure, that regards them as the same
variable, gives a different result. To make the point it suffices
to consider variation of the kinetic-gravitation part of the Lagrangian, which,
conveniently, has the same form (\ref{6.17}) in both Majorana and Standard models.

With help of the properties $<A>=<\tilde{A}>$ and $<AB>=<BA>$, we obtain
\begin{eqnarray}
\delta<h^{-1}g^\mu\partial_\mu\Psi i\gamma_0\tilde{\Psi}>
& = &2<h^{-1}g^\mu\partial_\mu \Psi i\gamma_0\delta\tilde{\Psi}>
+<\partial_\mu(h^{-1}g^\mu )\Psi i\gamma_0\delta\tilde{\Psi}>\nonumber\\
& & -\partial_\mu<h^{-1}g^\mu \Psi i\gamma_0\delta\tilde{\Psi}>\nonumber
\end{eqnarray}
and
\begin{eqnarray}
\delta<g^\mu\frac{1}{2}\omega_\mu\Psi i\gamma_0\tilde{\Psi}>
=\frac{1}{2}<(g^\mu \omega_\mu +\omega_\mu g^\mu)\Psi i\gamma_0\delta\tilde{\Psi}>.
\nonumber
\end{eqnarray}
These combine to give us
\begin{eqnarray}
\delta<h^{-1}D\Psi i\gamma_0\tilde{\Psi}>
=2h^{-1}<[g^\mu(\partial_\mu+\frac{1}{2}\omega_\mu)+ \frac{1}{2}hD_\mu(h^{-1}g^\mu)]
\Psi i\gamma_0\delta\tilde{\Psi}>
\label{8.18}
\end{eqnarray}
where
\begin{equation}
hD_\mu(h^{-1}g^\mu)=h\partial_\mu(h^{-1}g^\mu)+\omega_\mu\times g^\mu,
 \label{8.19}
\end{equation}
and we have dropped the boundary term with vanishing variation.

Completing the variation of other terms we find that the generalized Dirac equation
(\ref{8.9}) must be amended to the form
\begin{equation}
g^\mu\mathcal{D}_\mu\Psi i\gamma_3+\frac{1}{2}hD_\mu(h^{-1}g^\mu)\Psi i\gamma_3
=\Psi \underline{m}\,,\label{8.20}
\end{equation}
The additional term is proportional to the torsion, so it vanishes when the torsion
vanishes, but we have noted that is not generally the case.

In the usual way, the commutator of the generalized coderivative (\ref{8.6}) gives us
a generalized curvature:
\begin{equation}
[\mathcal{D}_\mu, \mathcal{D}_\nu] \Psi=\frac{1}{2} R(g_\mu\wedge g_\nu)\Psi
+\Psi W_{\mu\nu},\label{8.21}
\end{equation}
where $ R(g_\mu\wedge g_\nu)$ is the {\it gravitational curvature} (\ref{5.10}),
and the {\it electroweak curvature} has the standard form
\begin{eqnarray}
W_{\mu\nu}&=&i(\partial_\mu W_\nu-\partial_\nu W_\mu) -2W_\mu\times W_\nu\nonumber\\
&=&\frac{g}{2}i(-\partial_\mu \mathbf{A}_\nu+\partial_\nu \mathbf{A}_\mu
+2i\mathbf{A}_\mu\times \mathbf{A}_\nu)
+\frac{g'}{2}i(\partial_\mu B_\nu-\partial_\nu B_\mu)\label{8.22}.
\end{eqnarray}
Gravelectroweak theory obviously enhances the symmetry between these
two curvatures, and thus raises the question as to how far the symmetry should extend
throughout the theory.
Many people have noted the dyssymmetry in standard gauge field Lagrangians, which
are linear in the gravitational curvature but quadratic in the electroweak curvature.
Without trying to resolve the issue here, it is worth noting that GA
has been used in two studies of Lagrangians that are quadratic in the gravitational
curvature.\cite{Lewis06, Lasenby06b}

\section{Conclusions}\label{dis}

We have seen how STA enables integration of gravitational and electroweak interactions
into a unified {\it gravelectroweak gauge theory} with real spinor fields representing
basic physical entities.
Besides simplifying physical analysis and computations, as documented in the references,
this provides a new framework for studying interactions among gauge fields.

The most striking revelation of the new theory is that the real Dirac equation
has inherent geometric structure that perfectly matches the structure of electroweak
isospace, including precisely the right number of degrees of freedom.
This means that electroweak theory can be construed as a theory of spacetime geometry
of real spinor fields.

Incorporating the standard Weinberg-Salam model into this geometric framework reveals
a natural way to identify the chiral structure of the model with splitting of the gauge
fields into charged and neutral components. Though this insight cannot be expected to
yield new experimental predictions, it does open new possibilities for improving the
theory.  The most promising possibility has been dubbed the {\it Majorana model.}

We arrived at the Majorana model as a variant of the Standard model simply by a change
in projection operator.
{\it Chiral splitting} of the Dirac wave function by the operator $\frac{1}{2}(1 -\bsig_3)$
is motivated experimentally but not theoretically.
{\it Majorana splitting} by the operator $\frac{1}{2}(1 -\bsig_2)$ is
motivated theoretically but remains to be tested experimentally.

Since the  models are so similar structurally, it is not immediately clear
 whether the Majorana model is inconsistent with any current experimental
evidence supporting the Standard model.
But it is clear that experimental tests can be devised to discriminate between the models.

The crucial test of the Majorana model will be experimental validation of
zitterbewegung structure in the electron charge current. There are many possibilities
for experimental tests that cannot be addressed here. There are many metaphorical
remarks about zitterbewegung in the literature on relativistic QM. It is high time
to get serious about zitterbewegung as a real physical phenomenon.

There are other possibilities for experimental tests of the Majorana model, for
example, to distinguish Chiral and Majorana states in neutrino mixing.

Of course, if gravelectroweak theory works for leptons it probably works for
quarks as well. It remains to be seen if a comparable geometric theory can be
devised for strong interactions.

\vfill

\end{document}